\documentstyle[12pt]{article}

\newcommand{\be}{\begin{equation}}
\newcommand{\ee}{\end{equation}}
\newcommand{\ba}{\begin{array}}
\newcommand{\ea}{\end{array}}

\newtheorem{theorem}{Theorem}

\begin{document}

\title{\Large \bf Classification of Second Order Symmetric
         Tensors  in 5-Dimensional Kaluza-Klein-Type Theories}

\vspace{2mm}

\author{J. Santos \\
    {\it Universidade Federal do Rio G. do Norte} \\ {\em Departamento de
         F\'{\i}sica}, {\it Caixa Postal 1641}    \\
      {\em 59072-970 Natal - RN,  Brazil}   \\
     {\it E-mail\/}: janilo@dfte.ufrn.br  \\
                \\
         M.J. Rebou\c{c}as and A.F.F. Teixeira \\
    {\it Centro Brasileiro de Pesquisas F\'{\i}sicas}  \\
    {\em Departamento de Relatividade e Part\'{\i}culas} \\
    {\it Rua Dr Xavier Sigaud, 150}  \\
    {\it 22290-180 Rio de Janeiro - RJ,  Brazil}    \\
    {\it E-mail\/}: reboucas@cat.cbpf.br  }
\date{}

\maketitle

\begin{abstract}
An algebraic classification of second order symmetric tensors
in 5-dimensional Kaluza-Klein-type Lo\-rent\-zian spaces is
presented by using Jordan matrices.
We show that the possible Segre types are $[1,1111]$,
$[2111]$, $[311]$, $[z\,\bar{z}\,111]$, and the
degeneracies thereof. A set of canonical forms for each Segre
type is found. The possible continuous groups of symmetry
for each canonical form are also studied.
\end{abstract}

{\raggedright
\section{Introduction} }   \label{intro}
\setcounter{equation}{0}

It is well known that a coordinate-invariant characterization of
the gravitational field in general relativity is best given in
terms of the curvature tensor and a finite number of its covariant
derivatives relative to a canonically chosen field of Lorentz
frames~\cite{Cartan}~--~\cite{MacCSkea}.
The Riemann tensor itself is decomposable into three irreducible
parts, namely the Weyl tensor (denoted by $W_{abcd}$), the
traceless Ricci tensor ($S_{ab} \equiv  R_{ab} - \frac{1}{4}\,
R\, g_{ab}$) and the Ricci scalar ($R \equiv R_{ab}\, g^{ab}$).
The algebraic classification of the Weyl part of the
Riemann tensor, known as Petrov classification,  has played a
significant role in the study of various topics in general
relativity. However, for full classification of curvature tensor of
nonvacuum space-times one also has to consider the Ricci
part of the curvature tensor, which by virtue of Einstein's equations
$G_{ab} = \kappa\, T_{ab} + \Lambda \,g_{ab}\,$ clearly has the
same algebraic classification of both the Einstein tensor
$G_{ab} \equiv R_{ab} - \frac{1}{2}\, R\, g_{ab}\,$
and the energy-momentum tensor $T_{ab}$.

The algebraic classification of a sym\-metr\-ic two-\-tensor defined
on a four-\-di\-men\-sional Lorentzian manifold, known as
Segre classification, has been discussed by several authors~\cite{Hall}
and is of interest in at least three contexts. One is in understanding
some purely geometrical features of space-times (see, e.g., Churchill
\cite{Churc}, Pleba\'nski, \cite{Pleban} Cormack and Hall \cite{CorHal} ).
The second one is in classifying and interpreting matter field distributions~%
\cite{Hall1}~--~\cite{SanRebTei}.
The third is as part of the procedure for checking whether apparently
different space-times are in fact locally the same up to coordinate
transformations (equivalence problem
\cite{Karlh,MacCSkea}, \cite{Mac1}~--~\cite{MM}).

Over the past three decades there has been a resurgence in work
on Kaluza-Klein-type theories in five and more dimensions.
This has been motivated, on the one hand, by the quest for a
unification of gravity with the other fundamental interactions.
{}From a technical viewpoint, on the other hand, they have been
used as a way of finding new solutions of Einstein's equations
in four dimensions, without ascribing any physical meaning to
the additional components of the metric tensor~\cite{Gleiser}.

Using the theory of Jordan matrices we discuss, in this paper,
the algebraic classification of second order symmetric tensors
defined on five-dimensional (5-D for short) Lorentzian
manifolds $M$, extending previous results on this issue
\cite{Hall3}~--~\cite{Hall5}. We show that at a point $p \in M$
the Ricci tensor $R$ can be classified in four Segre types and their
twenty-two degeneracies. Using real half-null pentad bases for
the tangent vector space $T_{p}(M)$ to $M$ at $p$ we derive a set of
canonical forms for $R_{ab}$, generalizing the canonical
forms for a symmetric two-tensor on 3-D and 4-D space-times
manifolds \cite{Hall4,Hall3}. The continuous groups that
leave invariant each canonical form for $R_{ab}$ are also discussed.
Although the Ricci tensor is  constantly referred to,
the results of the following sections apply to any
second order real symmetric tensor on Kaluza-Klein-type
5-D Lorentzian manifolds.

{\raggedright
\section{Segre Types in 5-D Space-times}   }
\label{class}
\setcounter{equation}{0}

The essential idea underlying all classification is the concept of
equivalence. Clearly the objects may be grouped into different
classes according to the criteria one chooses to classify them.
In this section we shall classify $R^{a}_{\ b}$ up to similarity,
which is a criterion quite often used in mathematics and physics.
This approach splits the Ricci tensor in 5-D space-times into
four Segre types and their degeneracies.

Before proceeding to the classification of the Ricci tensor
let us state our general setting. Throughout this
paper $M$ is a 5-D space-time manifold endowed with a
Lorentzian metric $g$ of signature $(- + + + +),\;$
$T_{p}(M)$ denotes the tangent vector space to $M$ at a point
$p \in M$, and any latin indices but $p$ range
from 0 to 4.

Let $R_{ab}$ be the covariant components of a second order
symmetric tensor $R$ at $p \in M$. Given $R_{ab}$ we may
use the metric tensor to have the mixed form $R^{a}_{\ b}$ of
$R$ at $T_{p}(M)$. In this form the symmetric two-tensor
$R$ may be looked upon as a real linear operator
$R: T_{p}(M) \longrightarrow T_{p}(M)$. If one thinks
of $R$ as a matrix $R^{a}_{\ b}$, one can formulate the
eigenvalue problem
\be
R^{a}_{\ b}\, V^b = \lambda \, \delta^{a}_{\ b} \,V^{b}\,, \label{eigen}
\ee
where $\lambda$ is scalar and $V^b$ are the components of
a generic eigenvector $ {\bf V}  \in T_{p}(M)$.
The fact that we have $\delta^{a}_{\ b}$ rather than $g_{ab}$ on
the right hand side of equation (\ref{eigen}) makes apparent that
we have cast the non-standard eigenvalue problem involving the
hyperbolic (real) metric
$R_{ab}\,V^b = \lambda \,g_{ab}\,V^b$ into the standard form (\ref{eigen})
well known in linear algebra textbooks. However, we pay a price for
this because now $R^{a}_{\ b}$ is no longer symmetric in
general. We remind that in a space with positive (or negative)
definite metric a real symmetric operator can always be
diagonalized over the reals. Despite this problem we shall work with
the eigenvalue problem for $R^{a}_{\ b}$ in the standard
form (\ref{eigen}).

We first consider the cases when all eigenvalues of $R$ are real.
For these cases similarity transformations possibly exist \cite{Shilov}
under which $R^{a}_{\ b}$ takes at $p$ either one of the following
Jordan canonical forms (JCF for short):
\vspace{1mm}  \begin{sloppypar}
\[
\label{jmatrices}
\ba{ccc}
\; \left(
  \ba{ccccc}
  \lambda_{1} &    1        &    0         &    0        &    0      \\
       0      & \lambda_{1} &    1         &    0        &    0      \\
       0      &    0        & \lambda_{1}  &    1        &    0       \\
       0      &    0        &    0         & \lambda_{1} &    1       \\
       0      &    0        &    0         &    0        & \lambda_{1}
  \ea
  \right)\,,
&
\; \left(
  \ba{ccccc}
  \lambda_{1} &    1        &    0         &    0        &    0      \\
       0      & \lambda_{1} &    1         &    0        &    0      \\
       0      &    0        & \lambda_{1}  &    1        &    0       \\
       0      &    0        &    0         & \lambda_{1} &    0       \\
       0      &    0        &    0         &    0        & \lambda_{2}
  \ea
  \right)\,,
&
\; \left(
  \ba{ccccc}
  \lambda_{1} &    1        &    0         &    0        &    0      \\
       0      & \lambda_{1} &    1         &    0        &    0      \\
       0      &    0        & \lambda_{1}  &    0        &    0       \\
       0      &    0        &    0         & \lambda_{2} &    1       \\
       0      &    0        &    0         &    0        & \lambda_{2}
  \ea
  \right)\,,    \\
\mbox{(a) Segre type [5]} & \mbox{(b) Segre type [41]} &
                                              \mbox{(c) Segre type [32]}
\ea
\]  \end{sloppypar}
\vspace{1mm}  \begin{sloppypar}
\[
\ba{ccc}
\; \left(
  \ba{ccccc}
  \lambda_{1} &    1        &    0         &    0        &    0      \\
       0      & \lambda_{1} &    1         &    0        &    0      \\
       0      &    0        & \lambda_{1}  &    0        &    0       \\
       0      &    0        &    0         & \lambda_{2} &    0       \\
       0      &    0        &    0         &    0        & \lambda_{3}
  \ea
  \right)\,,
&
\; \left(
  \ba{ccccc}
  \lambda_{1} &    1        &    0         &    0        &    0      \\
       0      & \lambda_{1} &    0         &    0        &    0      \\
       0      &    0        & \lambda_{2}  &    1        &    0       \\
       0      &    0        &    0         & \lambda_{2} &    0       \\
       0      &    0        &    0         &    0        & \lambda_{3}
  \ea
  \right)\,,
&
\; \left(
  \ba{ccccc}
  \lambda_{1} &    1        &    0         &    0        &    0      \\
       0      & \lambda_{1} &    0         &    0        &    0      \\
       0      &    0        & \lambda_{2}  &    0        &    0       \\
       0      &    0        &    0         & \lambda_{3} &    0       \\
       0      &    0        &    0         &    0        & \lambda_{4}
  \ea
  \right)\,,   \\
\mbox{(d) Segre type [311]} & \mbox{(e) Segre type [221]} &
                                              \mbox{(f) Segre type [2111]}
\ea
\] \end{sloppypar}

\[
\ba{c}
\; \left(
  \ba{ccccc}
  \lambda_{1} &    0        &    0         &    0        &    0      \\
       0      & \lambda_{2} &    0         &    0        &    0      \\
       0      &    0        & \lambda_{3}  &    0        &    0       \\
       0      &    0        &    0         & \lambda_{4} &    0       \\
       0      &    0        &    0         &    0        & \lambda_{5}
  \ea
  \right) \,, \\
\mbox{(g) Segre type [1,1111]}  \vspace{1mm}
\ea
\]
or one of the possible block-degenerated Jordan matrices.
Here $\lambda_1, \cdots ,\lambda_5 \in {\cal R}$.
In the Jordan canonical form, a matrix consists of Jordan
blocks (Jordan submatrices) along the principal diagonal.
The above Segre types are nothing but their Segre characteristic,
well known in linear algebra and algebraic geometry. The Segre
characteristic is a list of digits inside square brackets, where
each digit refers to the multiplicity of the corresponding
eigenvalue, which clearly is equal to the dimension of the
corresponding Jordan block. A Jordan matrix $J$ is determined
from $R$ through similarity transformations $X^{-1} R\, X = J$,
and it is uniquely defined up to the ordering of the Jordan blocks.
Further, regardless of the dimension of a Jordan block
there is only one eigenvector associated to each block.
Degeneracy amongst eigenvalues in different Jordan
blocks will be indicated by enclosing the
corresponding digits inside round brackets. For each
degeneracy of this type it is easy to show that there exists
an invariant subspace of eigenvectors (eigenspace), whose dimension
is equal to the number of distinct Jordan submatrices with same
eigenvalue. Thus, e.g., in the above case (d) if
$\lambda_1 = \lambda_2$ then the Segre type is $[(31)1]$, and
besides the obvious one-dimensional invariant subspace defined
by an eigenvector associated to the eigenvalue $\lambda_3$,
there is a two-dimensional invariant subspace of eigenvectors
(2-eigenspace) with eigenvalue $\lambda_1$.
Finally, it is worth noting that the Segre type $[1,1111]$ and
its degeneracies are the only types that admit a timelike
eigenvector. We shall clarify this point in the next section.
The comma in these cases is used
to separate timelike from spacelike eigenvectors.

We shall show now that the Lorentzian
character of the metric $g$ on $M$, together with the
symmetry of $R_{ab}$, rule out the above cases
(a), (b), (c) and (e). Actually as the procedure
for eliminating all these cases are similar, and a proof
for the case (a) was already briefly outlined in Santos et al.
\cite{SRT}, for the sake of brevity we shall discuss here
in details how to eliminate the cases (c) and (e),
leaving to the reader to work out the details for the case (b).

{}From linear algebra we learn that to bring $R^a_{\ b}$
to a Jordan canonical $J^a_{\ b}$ form there must exist a
nonsingular matrix $X$ such that
\be
X^{-1}R\,X=J\,.   \label{simtran}
\ee

We shall first consider the above case (c), where $J$ is the
Jordan canonical matrix for the Segre type $[32]$. Multiplying
the matricial equation (\ref{simtran}) from the left by $X$ and
equating the columns of both sides of the resulting matricial
equation we have the Jordan chain relations
\begin{eqnarray}
R{\bf X_1} & = & \lambda _1 {\bf X_1}             \,,  \label{first} \\
R{\bf X_2} & = & \lambda _1 {\bf X_2} + {\bf X_1} \,,  \label{second} \\
R{\bf X_3} & = & \lambda _1 {\bf X_3} + {\bf X_2} \,,  \label{third} \\
R{\bf X_4} & = & \lambda _2 {\bf X_4}             \,,  \label{fourth} \\
R{\bf X_5} & = & \lambda _2 {\bf X_5} + {\bf X_4} \,,  \label{fifth}
\end{eqnarray}
where we have denoted by ${\bf X_A}$ (${\bf A} =1, \cdots ,5$) the column
vectors of the matrix $X$ and beared in mind that here the matrix $J$ is
that of case (c). As $R$ is a symmetric two-tensor, from equations
(\ref{first}) and ({\ref{second}) one easily obtains
\be
\lambda_1 {\bf X_1}.{\bf X_2} =
\lambda_1 {\bf X_2}.{\bf X_1} + {\bf X_1}.{\bf X_1} \,, \label{sixth}
\ee
where the scalar products defined by a Lorentzian metric $g$ are
indicated by a dot between two vectors. Equation (\ref{sixth})
implies that ${\bf X_1}$ is a null vector.
Similarly  eqs.\ (\ref{fourth}) and (\ref{fifth}) imply that
${\bf X_4}$ is also a null vector. Moreover, if
$\lambda_1 \not= \lambda_2$ from eqs.\ (\ref{first})
and (\ref{fourth}) one finds  that ${\bf X_1}.{\bf X_4} = 0$.
If $\lambda_1 = \lambda_2$ equations (\ref{first}), (\ref{fourth})
and (\ref{fifth}) imply again that ${\bf X_1}.{\bf X_4} = 0$.
In short, ${\bf X_1}$ and ${\bf X_4}$ are both null and
orthogonal to each other. As the metric $g$ on $M$ is locally
Lorentzian the null vectors ${\bf X_1}$ and ${\bf X_4}$
must be collinear, i.e., they are proportional. Hence $X$ is
a singular matrix, which contradicts our initial assumption
regardless of whether $\lambda_1 = \lambda_2$ or
$\lambda_1 \not= \lambda_2$. So, there is no nonsingular
matrix $X$ such that equation (\ref{simtran}) holds for
$J$ in the JCF given in case (c). In other words, at a point
$p \in M$ the Ricci tensor $R$ defined on 5-D Lorentzian
manifolds $M$ cannot be Segre types $[32]$ and its
degeneracy $[(32)]$.

Regarding the Segre type $[221]$ case, by analogous calculations
one can show that equation (\ref{simtran}) gives rise to the
following Jordan chain:
\begin{eqnarray}
R{\bf X_1} & = & \lambda_1 {\bf X_1}                \,,  \label{1st} \\
R{\bf X_2} & = & \lambda_1 {\bf X_2} + {\bf X_1}    \,,  \label{2nd} \\
R{\bf X_3} & = & \lambda_2 {\bf X_3}                \,,  \label{3rd} \\
R{\bf X_4} & = & \lambda_2 {\bf X_4} + {\bf X_3}    \,,  \label{4th} \\
R{\bf X_5} & = & \lambda_3 {\bf X_5}                \,.  \label{5th}
\end{eqnarray}
As the two pairs of equations (\ref{1st})--(\ref{2nd}) and
(\ref{4th})--(\ref{5th}) have the same algebraic structure
of the two pairs of equations (\ref{first})--(\ref{second}) and
(\ref{fourth})--(\ref{fifth}) of the case (c), they can similarly
be used to show that ${\bf X_1}$ and ${\bf X_3}$ are null vectors.
Moreover, here for any $\lambda_3$ and regardless of whether
$\lambda_1 = \lambda_2$ or $\lambda_1 \not= \lambda_2$ one can
show that ${\bf X_1}$ and ${\bf X_3}$ are  orthogonal to one
another. Thus, at a point $p \in M$, the Ricci tensor defined on
5-D Lorentzian manifolds cannot be brought to the Jordan canonical
forms corresponding to the Segre types $[221]$, $[(22)1]$, $[2(21)]$
and $[(221)]$.

\vspace{1mm}
In brief, for real eigenvalues we have been left solely with the
Segre type cases (d), (f) and (g) and their degeneracies as
JCF for $R^{a}_{\ b}$. In other words, when all eigenvalues
are real a second order symmetric tensor on a 5-D Lorentzian $M$
manifold can be of one of the following Segre types at $p \in M$:
\vspace{3mm}
\begin{enumerate}
\item \begin{sloppypar}
$[1,1111]$ and its degeneracies $[1,11(11)]$, $[(1,1)111]$,
$[1,(11)(11)]$, $[(1,1)(11)1]$, $[1,1(111)]$, $[(1,11)11]$,
$[(1,1)(111)]$, $[(1,11)(11)]$, $[1,(1111)]$, $[(1,111)1]$ and
$[(1,1111)]$;  \end{sloppypar}
\item
$[2111]$ and its specializations $[21(11)]$, $[(21)11]$, $[(21)(11)]$,
$[2(111)]$, $[(211)1]$ and $[(2111)]$;
\item
$[311]$ and its degeneracies $[3(11)]$, $[(31)1]$ and $[(311)]$.
\end{enumerate}
\vspace{2mm}
As a matter of fact, to complete the classification for the cases
where the characteristic equation corresponding to (\ref{eigen})
has only real roots one needs to show that the above remaining Segre
types are consistent with the symmetry of $R_{ab}$ and the
Lorentzian character of the metric $g$. In the next section we
will find a set of canonical forms of $R_{ab}$ corresponding
to the above Segre types, which makes apparent that
these types fulfill both conditions indeed.
\vspace{2mm}

In the remainder of this section we shall discuss the
cases when the $R^a_{\ b}$  has complex eigenvalues.
It is well known that if $z_1$ is complex root of a
polynomial with real coefficients so is its complex
conjugate $\bar{z}_1$. As the characteristic equation
associated to the eigenvalue problem (\ref{eigen}) is a
fifth order polynomial with real coefficients it must
have at least one and may have at most three real roots.
Accordingly, the characteristic polynomial will have at most four
and at least two complex roots. As far as the multiplicities
are concerned one can easily realize that while the
complex roots can be either single or double degenerated,
the real ones can have multiplicity 1, 2 and 3. Taking into
account these remarks it is easy to figure out that when
complex eigenvalues occur the possible Jordan matrices
$J^a_{\ b}$ for $R^a_{\ b}$ are
\vspace{1mm}
\[
\label{cjmatrices}
\ba{ccc}
\; \left(
  \ba{ccccc}
    z_{1}     &    0        &    0         &    0        &    0      \\
       0      & \bar{z}_{1} &    0         &    0        &    0      \\
       0      &    0        & \lambda_{1}  &    1        &    0       \\
       0      &    0        &    0         & \lambda_{1} &    1       \\
       0      &    0        &    0         &    0        & \lambda_{1}
  \ea
  \right)\,,
&
\; \left(
  \ba{ccccc}
     z_{1}    &    0        &    0         &    0        &    0      \\
       0      & \bar{z}_{1} &    0         &    0        &    0      \\
       0      &    0        & \lambda_{1}  &    1        &    0       \\
       0      &    0        &    0         & \lambda_{1} &    0       \\
       0      &    0        &    0         &    0        & \lambda_{2}
  \ea
  \right)\,,
&
\; \left(
  \ba{ccccc}
     z_{1}    &    0        &    0         &    0        &    0      \\
       0      & \bar{z}_{1} &    0         &    0        &    0      \\
       0      &    0        & \lambda_{1}  &    0        &    0       \\
       0      &    0        &    0         & \lambda_{2} &    0       \\
       0      &    0        &    0         &    0        & \lambda_{3}
  \ea
  \right)\,,    \\
  \mbox{(i) Segre type [$z\, \bar{z}\, 3$]}
& \mbox{(ii) Segre type [$z\, \bar{z}\, 21$]}
& \mbox{(iii) Segre type [$z\, \bar{z}\,111$]}
\ea
\]
\vspace{2mm}
\[
\ba{ccc}
\; \left(
  \ba{ccccc}
     z_{1}    &    0        &    0         &    0        &    0      \\
       0      & \bar{z}_{1} &    0         &    0        &    0      \\
       0      &    0        &  w_{1}       &    0        &    0       \\
       0      &    0        &    0         & \bar{w}_{1} &    0       \\
       0      &    0        &    0         &    0        & \lambda_{1}
  \ea
  \right)\,,
&
\; \left(
  \ba{ccccc}
     z_{1}    &    1        &    0         &    0        &    0      \\
       0      & z_{1}       &    0         &    0        &    0      \\
       0      &    0        & \bar{z}_{1}  &    0        &    0       \\
       0      &    0        &    0         & \bar{z}_{1} &    0       \\
       0      &    0        &    0         &    0        & \lambda_{1}
  \ea
  \right)\,,
&
\; \left(
  \ba{ccccc}
     z_{1}    &    1        &    0         &    0        &    0      \\
       0      &   z_{1}     &    0         &    0        &    0      \\
       0      &    0        & \bar{z}_{1}  &    1        &    0       \\
       0      &    0        &    0         & \bar{z}_{1} &    0       \\
       0      &    0        &    0         &    0        & \lambda_{1}
  \ea
  \right)\,,  \\
\mbox{(iv) Segre type [$z\, \bar{z}\, w\, \bar{w}\, 1$]}
& \mbox{(v) Segre type [$2z\: \bar{z}\, \bar{z}\, 1$]}
& \mbox{(vi) Segre type [$2z\: 2\bar{z}\: 1$]}  \vspace{3mm}
\ea
\]
or one of the possible block-degenerated Jordan matrices
with degeneracies amongst real eigenvalues. Here
$\lambda_1, \lambda_2, \lambda_3 \in {\cal R} $,
$ w_1, \bar{w}_1, z_1, \bar{z}_1 \in {\cal C} $
and  $z, \bar{z}$ as well as $w, \bar{w}$ denote
complex conjugate eigenvalues.

One might think at first sight that an analogous procedure
to that applied to eliminate Jordan matrices with real
eigenvalues could be used here again to rule out some of the
above Segre types. However it turns out that the method does
not work when there are complex eigenvalues because
the basic result that two (real) orthogonal null vectors are
necessarily proportional does not hold for the complex
vectors. Thus, for the above Segre type case (vi), e.g.,
one can derive from the corresponding Jordan chain that the
first and the third column vectors of the matrix $X$
are both null and orthogonal to each other. Nevertheless,
this does not imply that $X$ is a singular matrix, inasmuch
as ${\bf X_1}$ and ${\bf X_3}$ are not necessarily proportional.
Thus, we shall use instead an approach  similar to that
employed by Hall \cite{Hall,Hall5} when dealing with
the complex Segre types in the classification of the
Ricci tensor in 4-D space-times manifolds.

Suppose that the $R^a_{\ b}$ has a complex eigenvalue
$z_1= \alpha + i\, \beta$ ($\,\alpha, \beta \in {\cal R} \,,\,
\beta \neq 0$) associated to the eigenvector
${\bf V} = {\bf Y} + i\, {\bf Z}$, with
components $V^a = Y^a + i\, Z^a$ relative to a basis in which
$R^a_{\ b}$ is real. The eigenvalue equation
\be
R^{a}_{\ b}\, V^b = z_1 \, V^{a} \label{ceigen1}
\ee
implies
\be
R^{a}_{\ b}\, \bar{V}^b = \bar{z}_1 \, \bar{V}^{a}\,, \label{ceigen2}
\ee
where the eigenvalue $\bar{z}_1 = \alpha - i\, \beta$ and
where $\bar{V}^a = Y^a - i\, Z^a$ are the components of a second
eigenvector ${\bf \bar{V} }$. Since $R_{ab}$ is symmetric we have
$\bar{V}_a \,R^{a}_{\ b}\, V^b = V_a \,R^{a}_{\ b}\, \bar{V}^b$,
which together with eqs.\ (\ref{ceigen1}) and (\ref{ceigen2})
yield $\bar{V}_a \, V^a = 0$, hence
\be
Y_a \, Y^a + Z_a \, Z^a = 0\,.  \label{tplane}
\ee
{}From this last equation it follows that either one of the
vectors ${\bf Y}$ or ${\bf Z}$ is timelike (and the other
spacelike) or both are null and, since $\beta \neq 0$,
not proportional to each other. Regardless of whether
${\bf Y}$ and ${\bf Z}$ are both null vectors or one
timelike and the other spacelike, the real and the imaginary
part of (\ref{ceigen1}) give
\begin{eqnarray}
R^{a}_{\ b}\, Y^{b} & = & \alpha Y^{a} - \beta Z^{a}\,, \label{2plane1} \\
R^{a}_{\ b}\, Z^{b} & = & \beta Y^{a} + \alpha Z^{a}\,.   \label{2plane2}
\end{eqnarray}
Therefore, in either case the vectors ${\bf Y}$ and ${\bf Z}$ span
a timelike invariant 2-space of $T_{p}(M)$ under $R^{a}_{\ b}$.
The 3-space orthogonal to this timelike 2-space is spacelike
and must have three spacelike orthogonal eigenvectors
(${\bf x\,, y}$ and ${\bf z}$, say) of $R^{a}_{\ b}$ with real eigenvalues
(see next section for more details about this point).
These spacelike eigenvectors together with ${\bf V}$ and
${\bf \bar{V} }$ complete a set of five linearly
independent eigenvectors of $R^{a}_{\ b}$ at $p \in M$. Therefore,
when there exists a complex eigenvalue $R^{a}_{\ b}$ is necessarily
diagonalizable over the complex field and possesses three real eigenvalues.
In other words, among the possible JCF for $R^{a}_{\ b}$
with complex eigenvalues only that of case (iii) and
its degeneracies are allowed, i.e., only the Segre
type $[z\, \bar{z}\, 111]$ and its specializations
$[z\, \bar{z}\, 1(11)]$ and $[z\, \bar{z}\, (111)]$
are permitted.

We can summarize the results of the present section by
stating the following theorem:
\begin{theorem} \label{Stypes}
Let $M$ be a real five-dimensional manifold endowed
with a Lorentzian metric $g$ of signature {\rm(}$ - + + + +${\rm)}.
Let $R^a_{\ b}$ be the mixed form of a second order
symmetric tensor $R$ defined at any point $p \in M$. Then
$R^a_{\ b}$ takes one of the following Segre types:
$[1,1111]$, $[2111]$, $[311]$, $[z\,\bar{z}\,111]$, or some
degeneracy thereof.
\end{theorem}

{\raggedright
\section{Canonical Forms} }
\label{canon}
\setcounter{equation}{0}

In the previous section we have classified the Ricci tensor
up to similarity transformations. A further refinement to that
classification  is to choose exactly one element, as simple
as possible, from each class of equivalent objects. The
collection of all such samples constitutes a set of
canonical forms for $R^a_{\ b}$.
In this section we shall obtain such a set for the
symmetric Ricci tensor in terms of semi-null pentad bases
of vectors ${\cal B}= \{{\bf l},{\bf m},{\bf x},{\bf y},{\bf z}\}$,
whose non-vanishing inner products are
\be
l^{a}m_{a} = x^{a}x_{a} = y^{a}y_{a} = z^{a}z_{a} = 1. \label{inerp}
\ee

At a point $p \in M$ the set of second order symmetric tensors
on a 5-D manifold $M$ constitutes a 15-dimensional vector
space $\cal V$, which can be spanned by the following 15
basis symmetric tensors:
\begin{eqnarray}
&
l_a l_b\,,\;\; m_a m_b\,,\;\; x_a x_b\,,\;\; y_a y_b\,,\;\;
z_a z_b\,,\;\; 2\,l_{(a}m_{b)}\,,\;\; 2\,l_{(a}x_{b)}\,,\;\;
2\,l_{(a}y_{b)}\,,\;\; 2\,l_{(a}z_{b)}\,, & \nonumber \\
&
2\,m_{(a}x_{b)}\,,\;\; 2\,m_{(a}y_{b)}\,,\;\; 2\,m_{(a}z_{b)}\,,\; \;
2\,x_{(a}y_{b)}\,,\;\; 2\,x_{(a}z_{b)}\,,\;\; 2\,y_{(a}z_{b)}\,.&
\label{sybase}
\end{eqnarray}
Clearly any symmetric two-tensor at $p \in M$ is a vector in $\cal V$
and can be written as a linear combination of the basis elements
(\ref{sybase}). So, for example, bearing in mind the non-vanishing
inner products given by (\ref{inerp}), one can easily work out the
following decomposition (completeness relation) for the metric:
\be
g_{ab}=2\,l_{(a}m_{b)} + x_{a}x_{b} + y_{a}y_{b} + z_a z_b\,. \label{gab}
\ee
As far as the Ricci tensor is concerned the most
general decomposition of $R_{ab}$ in terms of the above
semi-null basis is manifestly given by
\begin{eqnarray}
R_{ab} & = &
2\,\rho_{1}\,l_{(a}m_{b)} + \rho_{2}\,l_{a}l_{b} + \rho_{3}\,x_{a}x_{b} +
\rho_{4}\,y_{a}y_{b}+ \rho_{5}\,z_{a}z_{b}+\rho_{6}\,\,m_{a}m_{b} \nonumber \\
&  &
+2\,\rho_{7}\,l_{(a}x_{b)} + 2\,\rho_{8}\,l_{(a}y_{b)} +
2\,\rho_{9}\,l_{(a}z_{b)} + 2\,\rho_{10}\,m_{(a}x_{b)} +
2\,\rho_{11}\,m_{(a}y_{b)}  \nonumber \\
&  &
+ 2\,\rho_{12}\,m_{(a}z_{b)} + 2\,\rho_{13}\,x_{(a}y_{b)} +
2\,\rho_{14}\,x_{(a}z_{b)} + 2\,\rho_{15}\,y_{(a}z_{b)}\,,
\label{rabgen}
\end{eqnarray}
where the coefficients $\rho_{1}, \ldots ,\rho_{15} \in {\cal R} $.

In the remainder of this section we shall show that
for each of the Segre types of the theorem \ref{Stypes}
a semi-null pentad basis with non-vanishing inner products
(\ref{inerp}) can be introduced at $p \in M$ such that
$R_{ab}$ takes one and only one of the following
canonical forms:
\begin{eqnarray}
\!\!\![1,1111] \: & R_{ab} = & 2\,\rho_1\,l_{(a}m_{b)} +
\rho_2\,(l_{a}l_{b} + m_{a}m_{b}) + \rho_3\,x_{a}x_{b} +
\rho_4\,y_{a}y_{b} + \rho_5\,z_{a}z_{b} \,,  \label{rab11111}  \\
\!\!\!{[}2111] \: & R_{ab} = & 2\,\rho_1\,l_{(a}m_{b)} \pm l_{a}l_{b} +
\rho_3\,x_{a}x_{b} + \rho_4\,y_{a}y_{b} + \rho_5\,z_{a}z_{b}\,,
                                                   \label{rab2111}  \\
\!\!\!{[}311] \: & R_{ab} = & 2\,\rho_1\,l_{(a}m_{b)} + 2l_{(a}x_{b)} +
\rho_1\,x_{a}x_{b} + \rho_4\,y_{a}y_{b} + \rho_5\,z_{a}z_{b}\,,
                                               \label{rab311} \\
\!\!\!{[}z\,\bar{z}\,111] \: & R_{ab} = & 2\,\rho_1\,l_{(a}m_{b)} +
\rho_2\,(l_{a}l_{b} - m_{a}m_{b}) +\rho_3\,x_{a}x_{b} +
\rho_4\,y_{a}y_{b} + \rho_5\,z_{a}z_{b}\,,
                                                \label{rabzz111}
\end{eqnarray}
where $\rho_1, \cdots ,\rho_5 \in {\cal R}$ and $\rho_2\, \neq 0$
in (\ref{rabzz111}).

Before proceeding to the individual analysis of the above
Segre types it is worth noticing that since $R_{ab}$ is
symmetric the condition
\be
g_{ac}\, R^c_{\ b} = g_{bc}\, R^c_{\ a}  \label{concond}
\ee
must hold for each possible Segre type, where the metric
tensor $g$ is assumed to have a Lorentzian signature
($- + + + +$).

{\bf Segre type [1,1111]}. For this type if one writes down a
general symmetric real matrix $g_{ab}$, uses eq.\ (\ref{concond})
and the corresponding Jordan matrix (case (g) of the previous
section) one obtains
\be
g_{ab} = \; \mbox{diag}\: ( \mu_1, \mu_2, \mu_3, \mu_4, \mu_5)\,,
                                                      \label{gd}
\ee
where $\mu_1, \cdots ,\mu_5 \in {\cal R}$. As $\mbox{det}\: g_{ab} < 0$,
then all $\mu_a \not= 0$ \ ($a = 1, \cdots ,5$). Moreover, at least
one $\mu_a < 0$. As a matter of fact, owing to the Lorentzian
signature $(- + + + +)$ of the metric one and only one $\mu_a$
is negative. On the other hand $R^a_{\ b}$ is real, symmetric
and has five different eigenvalues. So, bearing in mind that
the scalar product on $T_p(M)$ is defined by $g_{ab}$,
the associated eigenvectors are orthogonal to each other
and their norms are equal to $\mu_a \not= 0$. Thus they are
either spacelike or timelike. Since one $\mu_a < 0$, one
of the eigenvectors must be timelike. Then the others are
necessarily spacelike. Hence, without loss of generality
one can always choose a basis with five orthonormal vectors
$\bar{\cal{B}} =\{{\bf t}, {\bf w}, {\bf x}, {\bf y}, {\bf z}\} $
defined along the invariant directions
so that
\be
R_{ab} = -\tilde{\rho}_1\, t_a t_b + \tilde{\rho}_2\, w_a w_b +
\rho_3\, x_a x _b + \rho_4\, y_a y_b + \rho_5\, z_a z_b \,,\label{twxyz}
\ee
where $-t^a t_a= w^a w_a = x^{a}x_{a} = y^{a}y_{a} = z^{a}z_{a} = 1$.

It should be noticed that the above  form for the
Segre type $[1,1111]$ makes apparent that this type as well
as its degeneracies are consistent with both the symmetry
of $R_{ab}$ and the Lorentzian signature of the metric
tensor $g$.

Finally, we introduce two null vectors
${\bf l} = \frac{1}{\sqrt{2}} ({\bf t} + {\bf w})$ and
${\bf m} = \frac{1}{\sqrt{2}} ({\bf t} - {\bf w})$
to form a semi-null basis
${\cal B}= \{{\bf l},{\bf m},{\bf x},{\bf y},{\bf z}\}$,
in terms of which $R_{ab}$ takes the canonical form
(\ref{rab11111}), with
\be
\rho_1 = \frac{1}{2}( \tilde{\rho}_1 + \tilde{\rho}_2 )
\qquad\mbox{and}\qquad
\rho_2 = \frac{1}{2}( \tilde{\rho}_2 - \tilde{\rho}_1 )\,.
\ee

We remark that according to the form (\ref{twxyz})
this type may degenerate into eleven other types
in agreement with the previous section. Further, they
are all diagonalizable with five linearly independent real
eigenvectors, one of which is necessarily timelike.
In the non-degenerated case $[1,1111]$ a semi-null
pentad basis $\cal B$ is uniquely determined \cite{foot1}.
When degeneracies exist, though, the form (\ref{twxyz})
(or alternatively (\ref{rab11111})) is invariant under
appropriate continuous transformations of the pentad basis of
vectors. Hereafter, in dealing with the symmetries of
any  Segre types we shall consider only continuous
transformations of the pentad basis of vectors.
Thus, for example, in the type $[(1,11) 11]$
the vectors ${\bf t}, {\bf w}$ and ${\bf x}$ are
determined up to 3-dimensional Lorentz rotations
$SO(1,2)$ in the invariant 3-space of eigenvectors
which contains these vectors. Similarly, for the Segre
type $[1, 1(111)]$ the vectors ${\bf x}, {\bf y}$ and
${\bf z}$ can be fixed up to 3-D spatial rotations
$SO(3)$, whereas the type $[1,11(11)]$  allows
spatial rotations in the plane $({\bf y},{\bf z})$.
Finally, the invariance group for the type $[(1,1111)]$
is the full generalized Lorentz group.

\vspace{2mm}
{\bf Segre type [2111]}. If one writes down a general
symmetric matrix $g_{ab}$ and uses the symmetry condition
(\ref{concond}) and the corresponding Jordan matrix $J^a_{\ b}$
for this type (case (f) in section 2) one obtains \vspace{2mm}
\be
\label{gab2}
g_{ab}\, = \,
\left(
\ba{ccccc}
   0           & \epsilon &     0       &     0         &   0   \\
\epsilon       & \gamma   &     0       &     0         &   0   \\
   0           &   0      & \epsilon_1  &     0         &   0   \\
   0           &   0      &     0       & \epsilon_2    &   0   \\
   0           &   0      &     0       &   0           & \epsilon_3
\ea
\right)\,,  \vspace{2mm}
\ee
where $\gamma, \epsilon,\epsilon_1, \epsilon_2  ,\epsilon_3 \in {\cal R}$.
Again from the corresponding Jordan matrix it also follows that
the column vectors
$l^a = (1, 0, 0, 0, 0)$,  $\tilde{x}^a = (0, 0, 1, 0, 0)$,
$\tilde{y}^a = (0, 0, 0, 1, 0)$ and  $\tilde{z}^a = (0, 0, 0, 0, 1)$
are the only independent eigenvectors of $J^a_{\ b}$, and
have associated eigenvalues $\lambda_1\,, \lambda_3\,, \lambda_4$ and
$\lambda_5$, respectively. These four vectors
together with the column null vector
$m^a= (-\gamma/(2\epsilon^2), 1/\epsilon, 0, 0, 0)$ constitute a
basis of $T_p(M)$. Clearly the spacelike vectors $({\bf \tilde{x}},
{\bf \tilde{y}},{\bf \tilde{z}})$ can be suitably normalized to form
a semi-null pentad basis ${\cal B} = \{{\bf l},{\bf m},{\bf x},
{\bf y},{\bf z}\}$ such that both equations (\ref{inerp}) and
(\ref{gab}) hold. Using now that ${\bf l},{\bf x}, {\bf y}$ and
${\bf z}$ are eigenvectors, eq.\ (\ref{rabgen}) simplifies to
\be
R_{ab} = 2\,\rho_1\, l_{(a}m_{b)} + \rho_2\, l_{a}l_{b} +
      \rho_3\, x_a x_b + \rho_4\, y_a y_b + \rho_5\, z_a z_b\,,
\ee
where the condition $\rho_2 \not= 0$ must be imposed otherwise
${\bf m}$ would be a fifth linearly independent eigenvector. A
transformation (null rotation) of the form
$l^a \rightarrow \zeta l^a,\: m^a \rightarrow m^a/\zeta,\, \zeta
\in {\cal R}^{+} $ can be used to set $\rho_2 = 1$ if $\rho_2 > 0$, and
$\rho_2 = -1$ if $\rho_2 < 0 $, making apparent that
(\ref{rab2111}) are the canonical forms for the Segre
type $[2111]$.

It is worth noticing that the above derivation of
(\ref{rab2111}) and the canonical form itself make apparent
that this Segre type as well as its degeneracies are
consistent with the Lorentzian signature of the metric
tensor and the symmetry of $R_{ab}$.

{}From the canonical form (\ref{rab2111}) one learns that
this type may degenerate into the six types we have enumerated
in the previous section. Although for this type we only have
four invariant directions intrinsically defined by $R^a_{\ b}$,
for the non-degenerated case $[2111]$ a semi-null pentad basis
$\cal B$ used in (\ref{rab2111}) can be uniquely fixed \cite{foot1}.
However, when there are degeneracies the canonical form (\ref{rab2111})
allows some freedom in the choice of a semi-null pentad
basis, i.e., the canonical form is invariant under some
appropriate continuous transformation of the basis of vectors.
Thus, e.g., the type $[21(11)]$ permits
local rotational symmetry (LRS for short) in the plane
(${\bf y}$, ${\bf z}$). Similarly a local null rotation symmetry
(LNRS) is allowed in the degenerated types $[(21)11]$ and
$[(21)(11)]$. Clearly this latter type admits both local
isotropies LRS and LNRS. Similar assertions
can obviously be made about other specializations of
the type $[2111]$, we shall not discuss them all here for
the sake of brevity, though.

{\bf Segre type [311]}. This case can be treated similarly to
the Segre type $[2111]$. From the associated Jordan matrix
$J^a_{\ b}$ it follows that the column vectors $l^a=(1, 0,0,0,0)$
$\tilde{y}^a = (0,0,0,1,0)$ and $\tilde{z}^a=(0,0,0,0,1)$ are the only
independent eigenvectors of $J^a_{\ b}$ with eigenvalues $\lambda_1$,
$\lambda_2$ and $\lambda_3$, respectively. Further, the
restrictions upon $g_{ab}$ imposed by eq.\ (\ref{concond}) give
$l^al_a = l^a \tilde{y}_a = l^a \tilde{z}_a = 0 $.
So, ${\bf l}$ is a null vector and $({\bf \tilde{y}}$,
${\bf \tilde{z}})$ are spacelike vectors. One can suitably
normalize these spacelike vectors and then select a semi-null
pentad basis of $T_p(M)$ containing ${\bf l}$, the normalized vectors
(${\bf y}$, ${\bf z}$) and two other vectors ${\bf x}$ and
${\bf m}$ with the orthonormality relations given by equation
(\ref{inerp}). Obviously this semi-null basis satisfies the
completeness relation (\ref{gab}). Now the fact that
${\bf l}$, ${\bf y}$ and ${\bf z}$ are eigenvectors of
$R^a_{\ b}$ can be used to reduce the general
decomposition (\ref{rabgen}) to
\be
R_{ab} = 2\,\rho_1\, l_{(a}m_{b)} + 2\,\rho_7\, l_{(a}x_{b) }+
      \rho_2\, l_{a}l_{b} + \rho_3\, x_a x_b + \rho_4\, y_a y_b +
      \rho_5\, z_a z_b\,,    \label{rab3i}
\ee
where the condition $\rho_7 \not= 0$ must be imposed otherwise
${\bf x}$ would be a fourth linearly independent eigenvector.
Besides,  as any linear combination of the form
$\kappa_1\, {\bf l} + \kappa_2 {\bf m} + \kappa_3 {\bf x}$ (with
$\kappa_1, \kappa_2 ,\kappa_3 \in {\cal R}$ and $\kappa_2,\kappa_3 \not= 0$)
must not be an eigenvector one finds that $\rho_3 = \rho_1$.
The transformation
\begin{eqnarray}
x^a   & \rightarrow  & x^a + 2\, \xi l^a\,,  \\
m^a   & \rightarrow  & m^a - 2\,\xi x^a - 2\, \xi^2 l^a\,, \vspace{1mm}
\end{eqnarray}
where $\xi = - \rho_2/(4\rho_7)$, yields $\rho_2=0$. Finally
a transformation (null rotation) $l^a \rightarrow l^a/\rho_7,\;
m^a \rightarrow \rho_7\, m^a$ can now be used to set $\rho_7 = 1$,
therefore reducing (\ref{rab3i}) to the canonical form
(\ref{rab311}).

Here again it is worth noting that the canonical form
(\ref{rab311}) itself and the method used to find it
make clear that the Segre type $[311]$ and its degeneracies
are consistent with the Lorentzian signature of the metric
tensor and the symmetry of $R_{ab}$.

{}From the canonical form (\ref{rab311}) one obtains that
this Segre type gives rise to three degenerated types
in agreement with the section 2.
Here again for the non-degenerated type $[311]$
a semi-null pentad basis $\cal B$ used in
(\ref{rab311}) can be fixed \cite{foot1}, but when there are
degeneracies the canonical form is invariant under some
appropriate continuous transformation of the basis of vectors.
So, for example, the type $[3(11)]$ permits LRS,
the Segre type $[(31)1]$ admits LNRS, and the type
$[(311)]$ allows both LRS and LNRS.

{\bf Segre type $[{\bf z\,\bar{z}\,111]}$}. For this type,
according the previous section $R^a_{\ b}$ necessarily has
two complex eigenvectors ${\bf V} = {\bf Y} \pm i\, {\bf Z}$
with associated eigenvalues $ \alpha \, \pm i\, \beta$
($\,\alpha, \beta \in {\cal R} \,,\, \beta \neq 0$). Moreover, they
are orthogonal, i.e.\  (\ref{tplane}) holds, and the
real vectors ${\bf Y}$ and ${\bf Z}$ span a timelike
invariant 2-subspace of $T_p(M)$ under $R^a_{\ b}$, i.e.
eqs.\ (\ref{2plane1}) and (\ref{2plane2}) hold as well. This
timelike invariant plane contains two distinct null
directions, which we shall use to fix a pair of real null
vectors ${\bf l}$ and ${\bf m}$ of a semi-null
pentad basis. When ${\bf Y}$ and ${\bf Z}$ are both null one
can choose their directions to fix two suitably normalized
null vectors ${\bf l}$ and ${\bf m}$, necessary to form a
semi-null pentad basis.
When they are not null vectors one can, nevertheless,
use them to find out the needed two null vectors as
follows.

If ${\bf V} = {\bf Y} + i\, {\bf Z}$ is an eigenvector of $R^a_{\ b}$
so is ${\bf V'} = \rho \, e^{i\, \theta} ({\bf Y} + \,i {\bf Z})$\
whose components are obviously given by
\begin{eqnarray}
{Y'}^a & = &\rho\,\cos\theta\,Y^a - \rho\,\sin\theta\,Z^a \,,\label{Yprime}\\
{Z'}^a & = &\rho\, \cos\theta\,Z^a + \rho\,\sin\theta\, Y^a\,, \label{Zprime}
\end{eqnarray}
where $0 < \rho < \infty $ and $ 0 \le \theta < \pi$ \cite{foot3}.
Now, to form a semi-null basis we first need a pair of vectors
(${\bf Y'}$, ${\bf Z'}$) such that
\begin{eqnarray}
{Y'}^a \, {Y'}_a & = & {Z'}^a \, {Z'}_a = 0\,,    \label{YZnull} \\
{Y'}^a \, {Z'}_a & = & 1\,.                       \label{YZnotc}
\end{eqnarray}
Clearly from (\ref{tplane}) if ${\bf Y'}$ is a null
vector, so is ${\bf Z'}$. The substitution of (\ref{Yprime})
and (\ref{Zprime}) into (\ref{YZnull}) and (\ref{YZnotc})
gives rise to a pair of equations which can be
solved in terms of $\theta$  and $\rho$  to give
\begin{eqnarray}
\cot\,(2\theta) & = & \, \frac{Y^aZ_a}{Y^aY_a}\,,  \label{2the} \\
\rho^2   & = &  \frac{\sin\,(2\theta)}{Y^aY_a}\,, \label{rho2}
\end{eqnarray}
where $Y^aY_a \not= 0$, as ${\bf Y}$ is a non-null vector.
We then choose the two null vectors we need to form
a semi-null basis by putting ${\bf l} = {\bf Y'}$ and
${\bf m} = {\bf Z'}$. In terms of these vectors the
eigenvalue equation (\ref{ceigen1}) becomes
\be
R^{a}_{\ b}\,(l^b \pm i\, m^b) =               \label{einlm}
         (\alpha \pm i\, \beta) \,(l^a \pm i\, m^a) \,,
\ee
making clear that for this type one can always choose
a pair of null vectors such that $ l^a \pm i\, m^a$
are eigenvectors with eigenvalues  $\alpha \pm i\, \beta$.

Using now that $ l^a \pm i\, m^a$ are eigenvectors
eq. (\ref{rabgen}) simplifies to
\begin{eqnarray}
R_{ab} & = & 2\,\rho_{1}\,l_{(a}m_{b)} + \rho_{2}\,(l_{a}l_{b} - \,m_{a}m_{b})
+ \rho_{3}\,x_{a}x_{b} + \rho_{4}\,y_{a}y_{b}+
               \rho_{5}\,z_{a}z_{b}\nonumber \\
       &   &  + \,2\,\rho_{13}\,x_{(a}y_{b)}
              + 2\,\rho_{14}\,x_{(a}z_{b)}
              + 2\,\rho_{15}\,y_{(a}z_{b)}\,, \label{rabzi}
\end{eqnarray}
where $\rho_1 = \alpha$ and $\rho_2 = \beta \not= 0$.
Clearly using eqs.\ (\ref{gab}) and (\ref{rabzi})
one finds that the corresponding mixed matrix
$R^a_{\ b}$ takes a block diagonal form with two blocks.
The first one $T^a_{\ b}$ is a $(2 \times 2)$ matrix
acting on the timelike invariant 2-space. According
to (\ref{einlm}) it is diagonalizable over the complex field
${\cal C}$, with eigenvectors $l^a \pm i\, m^a$ and associated
eigenvalues $\alpha \pm i\, \beta$. The second block
$E^a_{\ b}$ is a $(3 \times 3)$ symmetric matrix acting
on the 3-space orthogonal to the above timelike invariant
2-space. Hence it can be diagonalized by spatial rotation of
the basis vectors $({\bf x, y}$ and ${\bf z})$, i.e.,
through real similarity transformations. Thus,
there exists an orthogonal basis $({\bf \tilde{x}, \tilde{y}, \tilde{z}})$
relative to which $E^a_{\ b}$ takes a diagonal form with real
coefficients. In the basis $\bar{{\cal B}} =
\{{\bf l},{\bf m},{\bf \tilde{x}}, {\bf \tilde{y}},{\bf \tilde{z}}\}$
one must have
\be
R_{ab} =  2\,\alpha \,l_{(a}m_{b)} +
\beta\,(l_{a}l_{b} -m_{a}m_{b}) + \tilde{\rho}_3\,\tilde{x}_{a}\tilde{x}_{b} +
\tilde{\rho}_4\,\tilde{y}_{a}\tilde{y}_{b} +
\tilde{\rho}_5\,\tilde{z}_{a}\tilde{z}_{b}\,, \label{rabzziii}
\ee
rendering explicit that there are three spacelike orthogonal
eigenvectors of $R^a_{\ b}$ with real eigenvalues, as we have
pointed out in the previous section. Actually, the above
three orthogonal spacelike eigenvectors
(${\bf \tilde{x}}$, ${\bf \tilde{y}}$, ${\bf \tilde{z}}$)
constitute a basis of the 3-space (3-eigenspace) orthogonal to
the invariant timelike 2-subspace (2-eigenspace) of $T_p(M)$
spanned by either (${\bf l}, {\bf m}$) or (${\bf Y}$, ${\bf Z}$),
in agreement with the section 2.

Finally, one can  suitably normalize the above spacelike basis
vectors (${\bf \tilde{x}}$, ${\bf \tilde{y}}$, ${\bf \tilde{z}}$),
then select a semi-null pentad basis of $T_p(M)$
containing  the above null vectors (${\bf l}$, ${\bf m}$) and
the normalized new spacelike vectors (${\bf x}$, ${\bf y}$, ${\bf z}$)
with the orthonormality relations given by equation
(\ref{inerp}). Thus, in terms of this semi-null pentad basis
$R_{ab}$ takes the canonical form (\ref{rabzz111}) for
this Segre type.

It should be noticed that the canonical form (\ref{rabzz111})
evince that this Segre type and its degeneracies
are consistent with the symmetry (\ref{concond}) and the
Lorentzian signature of the metric tensor $g$. Moreover,
eq.\ (\ref{rabzz111}) gives rise to two degeneracies,
namely $[z\,\bar{z}\,1(11)]$ and $[z\,\bar{z}\,(111)]$.

Here again for the non-degenerated case $[z\,\bar{z}\,111]$
a semi-null pentad basis $\cal B$ used in
(\ref{rabzz111}) can be fixed \cite{foot1}, but when there are
degeneracies the canonical form is invariant under
some appropriate continuous transformation of the basis of vectors.
So for the type $[z\,\bar{z}\,(111)]$ the vectors
${\bf x}$, ${\bf y}$ and ${\bf z}$ can only be fixed
up to 3-D spatial rotations $SO(3)$, whereas the type
$[z\,\bar{z}\,1(11)]$ allows LRS.

It should be stressed that the classification we have
discussed in this work applies to any second
order symmetric tensor $R$ at a point $p \in M$ and
can vary as $p$ changes in $M$.

To conclude, we should like to mention that the
classification and the canonical forms we have
studied in the present work generalize those discussed
by Graham Hall and colaborators for symmetric
two-tensors on 4-D and 3-D  space-time
manifolds \cite{Hall3,Hall4}.

{\raggedright
\section{Acknowledgments}  }
Special thanks are due to our friend Graham S. Hall for many
motivating and fruitful discussions held during his visit to
our Institute.
J. Santos is grateful to CAPES for the grant of a fellowship.


\vspace{7mm}

\begin{thebibliography}{99}

\bibitem{Cartan} E. Cartan, ``Le\c{c}ons sur la G\'{e}om\'{e}trie des
\'{E}spaces de Riemann'', Gauthier-Villars, Paris (1951). Reprinted,
\'Edi\-tions Jac\-ques Ga\-bay, Pa\-ris, 1988. English translation by J.
Glazebrook, Math.\ Sci.\ Press, Brookline (1983).

\bibitem{Karlh} A. Karlhede, {\em Gen.\ Rel.\ Grav.\/} {\bf 12},
693 (1980).

\bibitem{MacCSkea}
See M. A. H. MacCallum and J. E. F. Skea, ``{\sc sheep}:
A Computer Algebra System for General Relativity'', in {\em Algebraic
Computing in General Relativity, Lecture Notes from the First Brazilian
School on Computer Algebra\/}, Vol.\ II, edited by M. J. Rebou\c{c}as and
W. L. Roque.  Oxford U. P., Oxford (1994); and also the fairly extensive
literature therein quoted on the equivalence problem.

\bibitem{Hall} G. S. Hall, {\em Diff.\ Geom.\/} {\bf 12}, 53 (1984).
This reference contains an extensive bibliography on the classification
of the Ricci tensor on 4-dimensional Lorentzian space-times manifolds.

\bibitem{Churc} R. V. Churchill, {\em Trans.\ Amer.\ Math.\ Soc.\/}
{\bf 34}, 784 (1932).

\bibitem{Pleban} J. Pleba\'nski, {\em Acta Phys.\ Pol.\/} {\bf 26},
963 (1964).

\bibitem{CorHal} W. J. Cormack and G. S. Hall, {\em J. Phys.\ A\/}
{\bf 12}, 55 (1979).

\bibitem{Hall1} G. S. Hall, {\em Arab.\ J. Sci.\ Eng.\/} {\bf 9},
87 (1984).

\bibitem{Hall2} G. S. Hall and D. A. Negm, {\em Int.\ J. Theor.\ Phys.\/}
{\bf 25}, 405 (1986).

\bibitem{RebAmaTei} M. J. Rebou\c{c}as, J. E. {\AA}man and
A. F. F. Teixeira, {\em J.\  Math.\ Phys.\/} {\bf 27}, 1370 (1986).

\bibitem{RebAma} M. J. Rebou\c{c}as and J. E. {\AA}man,
{\em J.\  Math.\ Phys.\/} {\bf 28}, 888 (1987).

\bibitem{CRTS} M. O. Calv\~{a}o, M. J. Rebou\c{c}as,
A. F. F. Teixeira and W. M. Silva Jr., {\em J.\ Math.\
Phys.\/} {\bf 29}, 1127 (1988).

\bibitem{FMM} J. J. Ferrando, J.A. Morales and M. Portilla,
{\em Gen.\ Rel.\ Grav.\/} {\bf 22}, 1021 (1990).

\bibitem{RebTei} M. J. Rebou\c{c}as and A. F. F. Teixeira,
{\em J. Math.\ Phys.\/} {\bf 32}, 1861 (1991).

\bibitem{RebTei1} M. J. Rebou\c{c}as and A. F. F. Teixeira,
{\em J. Math.\ Phys.\/} {\bf 33}, 2855 (1992).

\bibitem{SanRebTei} J. Santos, M. J. Rebou\c{c}as and A. F. F. Teixeira,
{\em J. Math.\ Phys.\/} {\bf 34}, 186 (1993).

\bibitem{Mac1} M. A. H. MacCallum, ``Classifying Metrics in Theory and
 Practice'', in {\em Unified Field Theory in More Than 4 Dimensions,
 Including Exact Solutions}, edited by V. de Sabbata and E. Schmutzer.
 World Scientific, Singapore (1983).

\bibitem{Mac2} M. A. H. MacCallum, ``Algebraic Computing in General
 Relativity'', in {\em Classical General Relativity}, edited by W. B. Bonnor,
 J. N. Islam and M. A. H. MacCallum. Cambridge U. P., Cambridge (1984).

\bibitem{MM} M. A. H. MacCallum, ``Computer-aided Classification of
Exact Solutions in General Relativity'', in {\em General Relativity
and Gravitational Physics (9th Italian Conference)\/}, edited by
R. Cianci, R. de Ritis, M. Francaviglia, G. Marmo, C. Rubano and
P. Scudellaro. World Scientific Publishing Co., Singapore (1991).

\bibitem{Gleiser}
See, for example, R. J. Gleiser and M. C. Diaz,
{\it Phys.\ Rev.\ D \/} {\bf 37}, 3761 (1988) and
references therein quoted on this subject.

\bibitem{Hall4} G. S. Hall, T. Morgan and Z. Perj\'es,
{\it Gen.\ Rel.\ Grav.\/} {\bf 19}, 1137 (1987).

\bibitem{Hall3} G. S. Hall, {\em J. Phys.\ A\/} {\bf 9}, 541 (1976).

\bibitem{Hall5} G. S. Hall, ``Physical and Geometrical
Classification in General Relativity'', Brazilian Center for
Physics Research Monograph, CBPF-MO-001/93 (1993).

\bibitem{Shilov}
G.E. Shilov, {\it Linear Algebra\/} (Dover Publ. Inc., New York, 1977).

\bibitem{SRT}
J. Santos, M. J. Rebou\c cas and A. F. F. Teixeira, ``Classification
of the Ricci Tensor in 5-Dimensional Space-times,'' in
{\em Gravitation: the Spacetime Structure\/}, proceeding of the
``8th  Latin American Symposium on Relativity and Gravitation,'' edited
by P. Letelier and W.A. Rodrigues. World Scientific Publishing Co.,
Singapore (1994).

\bibitem{foot1} Actually the pentad basis can be fixed up to discrete
transformations, as inversions and relabelling of the basis vectors.

\bibitem{foot3} It should be noticed that the values of $\theta$
outside this range, i.e. $\theta \in [\pi, 2\,\pi)$, simply revert
the directions of ${\bf Y'}$ and ${\bf Z'}$. Thus, they need not
to be considered.

\end{thebibliography}
\end{document}